# Low-coherent optical diffraction tomography by angle-scanning illumination

KYEOREH LEE,[1,2] SEUNGWOO SHIN,[1,2] ZAHID YAQOOB,[3] PETER T. C. SO,[3,4,5,7] AND YONGKEUN PARK[1,2,6,*]

[1]*Department of Physics, Korea Advanced Institute of Science and Technology (KAIST), Daejeon 34141, Republic of Korea*
[2]*KAIST Institute for Health Science and Technology, Daejeon 34141, Republic of Korea*
[3]*Laser Biomedical Research Center, G. R. Harrison Spectroscopy Laboratory, Massachusetts Institute of Technology (MIT), Cambridge, Massachusetts 02139, USA*
[4]*Department of Mechanical Engineering, MIT, Cambridge, Massachusetts 02139, USA*
[5]*Department of Biological Engineering, MIT, Cambridge, Massachusetts 02139, USA*
[6]*Tomocube Inc., Daejeon 34051, Republic of Korea*
[7]*ptso@mit.edu*
*\*yk.park@kaist.ac.kr*

**Abstract:** We propose and experimentally demonstrate temporally low-coherent optical diffraction tomography (ODT) based on angle-scanning Mach–Zehnder interferometry. Using a digital micromirror device based on diffractive tilting, we successfully maintain full-field interference of incoherent light during every scan sequence. The ODT reconstruction principles for temporally incoherent illuminations are thoroughly reviewed and developed. Several limitations of incoherent illumination are also discussed, such as the nondispersive assumption, optical sectioning capacity, and illumination angle limitation. Using the proposed setup and reconstruction algorithms, we successfully demonstrate low-coherent ODT imaging of microspheres, human red blood cells, and eukaryotic cells.

## 1. Introduction

Quantitative phase imaging (QPI) is an emerging bio-imaging technique [1-3]. The advantage of QPI is its ability to provide two- and three-dimensional (3-D) structural and functional information in a label-free fashion, which enables observation of biological samples in their optimal conditions. Therefore, QPI has been utilized for long-term behavioral observation of biological samples and rapid quantitative analysis of biological specimens [4, 5]. QPI measures the intact complex amplitude of the scattered optical field, which is an important base unit for advanced modalities such as light scattering analysis [6-8], scattering parameters retrieval [9, 10], Jones matrix measurement [11, 12], synthetic aperture imaging [13-15], and optical diffraction tomography (ODT) [16-18]. The versatile character of QPI has enriched its usability.

Among the QPI approaches, 3-D QPI techniques such as ODT provide the 3-D refractive index (RI) distribution of a sample from the measured optical field at various illumination angles [19-23] or sample rotation [24-27]. Owing to its label-free nature, ODT enables visualizing 3-D cellular morphology as well as dynamics of intracellular organelles without any constraints. ODT has been actively utilized for various applications, including biophysics [28, 29], hematology [30], immunology [31], pharmacology [32], developmental biology [33], and nanotechnology [34].

Since the QPI is an interferometric microscopy technique, the majority of early QPI methods utilized coherent light sources (i.e., temporally coherent lasers) [35-38]. However, it soon turned out that high coherence disturbs the robust reconstruction of optical fields due to inevitable multiple scattering in an imaging system [39], also known as "coherent noise" or "speckle noise." Even for ideal optical setups, for instance, the sample coverslip interface always induces multiple scattering, and the parasitic fringes would be found in interferograms. Though several methods have been suggested to remove such noise [40-46], they frequently fail due to time-varying noise originating from sources such as system vibration, sample stage translation, objective lens drift, and light source frequency shifting.

Incoherent QPI techniques have been introduced to fundamentally address the issue. One obvious and general solution involves matching the optical path length (OPL) of the sample and the reference arms in typical interferometric setups [47-49]. Once the interference is maintained, the reconstruction sequences become identical to their coherent counterparts. Later, several "common-path" setups have been proposed to reduce the complexity of the interferometric setup and increase stability [50-52]. However, common-path techniques usually sacrifice generality by introducing assumptions or approximations regarding the sample or incident light, which may induce imaging artifacts [53-55].

Unlike incoherent QPI, performing incoherent ODT is not a simple task. The main difficulty here is the loss of coherence when a sample is illuminated at oblique angles [56]. Since the ODT requires optical field measurements at multiple angles, such decoherence is inevitable in common Mach–Zehnder (MZ) interferometer-based ODT setups. Although some incoherent ODT methods have been presented based on common-path geometries [22, 57-59], a more general incoherent ODT technique is still in demand.

Here, we report an incoherent angle-scanning ODT setup based on a classical MZ setup. We maintain interference in oblique angles by diffractive modulation of incident light using a digital micromirror device (DMD). We also discuss the principles of incoherent ODT reconstruction in detail based on previous literature [57, 58, 60]. Using the proposed setup and principles, we experimentally perform low-coherent ODT imaging of various samples.

## 2. Experimental setup

The optical setup is shown in Fig. 1. We used a commercial microscope body (IX71, Olympus Inc.) with objective (UPLSAPO 60XW, 60×, NA = 1.2, Olympus Inc.) and condenser lenses (UPLSAPO 60XW, 60×, $NA^{(i)}$ = 1.2, Olympus Inc.), where NA and $NA^{(i)}$ are the numerical apertures of the objective and condenser lenses, respectively. Here, one should notice that spatial coherence is important when specifying a single illumination angle. In this work, a supercontinuum source (EXR-4, NKT Photonics Inc.) was used as the spatially coherent broadband light source. An additional bandpass filter was used to minimize several issues that arise as the source bandwidth increases (see Section 4.1 for details) [Fig. 1(a)]. The OPLs of the two arms were matched with a translation stage in the reference arm. For image acquisition, we used a commercial monochromatic camera (MD120MU-SY, XIMEA GmbH) synchronized with the illumination unit. To maintain interference at every oblique illumination angle, major alterations were applied to the conventional MZ interferometry based ODT setup in the reference arm and illumination portion.

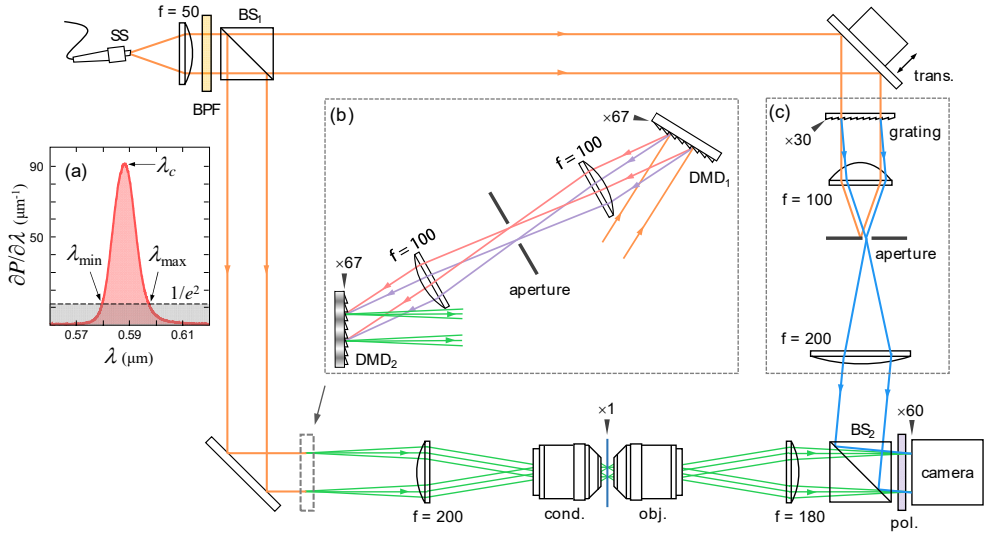

Fig. 1. Proposed optical setup. (a) Temporal illumination spectrum. The minimum and maximum wavelengths, $\lambda_{min}$ = 580 nm and $\lambda_{max}$ = 597 nm are defined by the $1/e^2$ bandwidth around the peak wavelength, $\lambda_c$ = 588 nm. (b) Light illumination unit composed of two DMDs. (c) Diffractive modulation based on off-axis scheme. The black arrows denote the sample conjugated planes with corresponding magnification factors. The denoted focal lengths of the lenses are in millimeters. The definitions of the abbreviations are as follows: SS, supercontinuum source; BPF, band-pass filter; BS, beam splitter; trans., translation stage; cond., condenser lens; obj., objective lens; and pol., polarizer.

### 2.1 Reference beam

In many QPI techniques, off-axis (or spatial modulation) schemes are widely used to obtain the optical field information in a single wide-field acquisition [37, 61]. For ODT techniques that require multiple optical fields, such utility becomes important for preventing motion of the sample during the entire series of measurements. However, as shown in Ref. [56], the off-axis configuration with incoherent light usually causes significant coherence loss. This is due to the dispersive nature of mirror-based spatial modulation, $2\pi \sin\theta_r / \lambda$, where $\lambda$ is the wavelength of light [Fig. 2(a)] and $\theta_r$ is the tilted angle of mirror. The dephasing between phasors $\exp(i2\pi x \sin\theta_r / \lambda)$ of different wavelengths becomes severe as the lateral position $x$ increases. Eventually, interference is diminished for $x\sin\theta_r > l_c$, where $l_c$ is the coherence length.

To prevent such decoherence, we introduce a diffraction grating (GT13-03, Thorlabs Inc.) in the reference arm [Fig. 1(c)]. Since the diffraction orders originate from the periodic structure of the grating, the $m$-th order diffraction peak exhibits identical spatial frequency regardless of wavelength, $2\pi N/\Lambda$, where $\Lambda$ is the period of the grating. For instance, the phasor becomes wavelength-independent when $N$ = 1 is selected. Therefore, clear interference at any lateral position $x$ could be observed, as in coherent situations [Fig. 2(b)]. We define such wavelength independent spatial modulation method as "diffractive modulation," which should be distinguished from conventional "reflective modulation" based on a mirror. Notice that introducing a diffraction grating in the reference beam generation is not a new concept [48, 49, 51, 62].

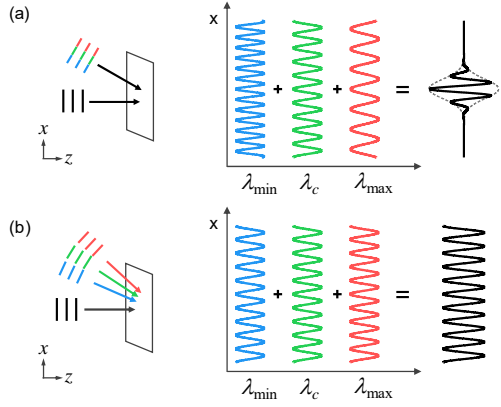

Fig. 2. Incoherent interference patterns in (a) reflective modulation, and (b) diffractive modulation. Red, greed, and blue colors denotes $\lambda_{max}$, $\lambda_c$, and $\lambda_{min}$, respectively. Severe decoherence is observed with reflective modulation due to the uneven periods of each phasor.

## 2.2 Illumination unit

We extend the diffractive modulation concept to illumination with incoherent light. Similar to the reference arm, reflection-based angle scanning (e.g., galvo mirrors) is no longer suitable in the sample arm of the proposed incoherent ODT. Rather, we require use of controllable diffractive elements such as spatial light modulators (SLMs). However, common liquid crystal on silicon (LCoS)-based SLMs usually have rather slow modulation speed and are not an ideal choice for ODT applications [59, 63]. Taking the modulation speed into account, we opted to use a DMD (DLP® LightCrafter™ 6500, Texas Instruments Inc.) for the illumination unit.

Despite the advantage of modulation speed, the DMD has not been a good option for broadband light because its intrinsic echelle grating geometry, which induces significant chromatic dispersion [64]. For a blazed grating with $\theta_b$ blaze angle, the brightest diffraction order ($N_{\bar{\uparrow}}$) is a function of wavelength,

$$N_{\bar{\uparrow}} = \text{nint}\left(\frac{\sin 2\theta_b}{\lambda} p_\parallel\right), \tag{1}$$

where $p_\parallel$ = 5.346 μm is the spatial period of the DMD parallel to the blaze direction (45° or diagonal), and nint(x) presents the integer nearest to x. Unlike transmission gratings that typically have small $\theta_b$, the large blaze angle of the DMD ($\theta_b = 12°$) induces $N_{\bar{\uparrow}}$ variation as a function of wavelength. According to Eq. (1), we find $N_{\bar{\uparrow}}$ = 4 for the DMD at a center wavelength $\lambda_c$ = 588 nm.

In coherent systems, the DMD blaze angle has commonly been compensated by corresponding oblique angle illumination [65, 66]. In incoherent systems, unfortunately, such reflective modulation is not permitted, as discussed above. In this report, we compensate such higher order diffraction and dispersive effects by introducing an identical DMD of idle state [DMD$_1$, Fig. 1(b)]. Since the configuration is symmetric to the light propagation direction, diffractions from the DMDs are completely undone with respect to the reciprocity of the setup. Therefore, the composite illumination unit becomes a binary amplitude (0 and 1) modulator without a blaze angle and is the desired non-dispersive diffractive modulator.

Based on our previous study using DMD, we used structured illumination with time multiplexing [66]. We compound four independent binary illuminations with the factors of power of 2 to effectively construct 4-bit structured illumination. Furthermore, we decompose the three plane waves in a cosine pattern, and we take four identical cosine patterns with different relative phases $\varphi$ = 0, $1\pi/4$, $2\pi/4$, and $3\pi/4$. Since we used 15 circular scanning cosine patterns (i.e., 30 oblique angle illuminations) and normal illumination, binary illumination was applied a total of 241 (=15 × 4 × 4 + 1) times. We adjusted the illumination NA ≈ 1.06 at $\lambda_c$, but the effective illumination NA is a function of wavelength (see Section 3.4 for detail). The entire series of binary patterns could be uploaded onto the DMD boards, which enables the DMD to be operated at maximum modulation speed (9,523 Hz). Unfortunately, the maximum speed could not be realized in this report due to the relatively slow acquisition speed of the camera.

## 3. ODT reconstruction from the incoherent QPI

The principle of ODT is well established from the general electromagnetic wave equations [67, 68]. In coherent systems, ODT reconstruction is the simple realization of the formulas from diffraction theory [16]. However, in incoherent cases, several fundamental and practical issues arise.

*3.1 Incoherent field measurement*

The incoherent system can be considered as a linear combination of coherent systems with different wavelengths. Therefore, the optical field retrieved from the incoherent interferogram can be described as

$$U(x,y) = U^{(i)}(x,y) + \int U_\lambda^{(s)}(x,y) \frac{\partial P}{\partial \lambda} d\lambda, \qquad (2)$$

where $U^{(i)}(x,y)$ is the incident plane wave $\exp\left[i\left(k_x^{(i)}x + k_y^{(i)}y\right)\right]$, which is independent of wavelength in the diffractive modulation geometry; $U_\lambda^{(s)}(x,y)$ is the scattered field as a function of wavelength $\lambda$. $\partial P/\partial \lambda$ is the normalized power spectral density of the light source, $\int \partial P/\partial \lambda \, d\lambda = 1$ [Fig. 1(a)].

At this point, the conventional coherent ODT reconstruction theorem could be applied to connect the monochromatic scattered field $\tilde{U}_\lambda^{(s)}(k_x, k_y)$ with unitless scattering potential $\chi_\lambda(x,y,z) = \left[n_\lambda(x,y,z)/m_\lambda\right]^2 - 1$ as

$$\tilde{U}_\lambda^{(s)}(k_x, k_y) = \frac{ik^2}{2k_z} \tilde{\chi}_\lambda\left(\mathbf{k} - \mathbf{k}^{(i)}\right), \qquad (3)$$

where $n_\lambda(x,y,z)$ is the 3-D RI distribution of a sample, $m_\lambda$ is the RI of a surrounding medium, $k = 2\pi m_\lambda/\lambda$ is the wavenumber of light, and $\mathbf{k}^{(i)} = \left(k_x^{(i)}, k_y^{(i)}, k_z^{(i)}\right)$ and $\mathbf{k} = \left(k_x, k_y, k_z\right)$ are the wavevectors of the incident and scattered field, respectively, that satisfy $k = |\mathbf{k}| = |\mathbf{k}^{(i)}|$ [67, 68]. By substituting Eq. (3) into Eq. (2), we can now associate incoherent QPI measurements with the sample scattering potential,

$$\tilde{U}^{(s)}(k_x, k_y) \equiv \tilde{U}(k_x, k_y) - \tilde{U}^{(i)}(k_x, k_y) = \int \frac{ik^2}{2k_z} \tilde{\chi}_\lambda\left(\mathbf{k} - \mathbf{k}^{(i)}\right) \frac{\partial P}{\partial k} dk. \qquad (4)$$

However, because each wavelength has its own equation [Eq. (3)], and no general relation holds between different $\tilde{U}_\lambda^{(s)}(k_x, k_y)$ or $\tilde{\chi}_\lambda\left(\mathbf{k} - \mathbf{k}^{(i)}\right)$, Eq. (4) is an ill-posed problem with an infinite number of possible solutions for a single measurement of $\tilde{U}^{(s)}$. Therefore, to reconstruct an ODT from incoherent field measurements, it is indispensable that we introduce proper assumptions to specify each $\tilde{U}_\lambda^{(s)}$. Here, similar to Ref. [57], we introduce the nondispersive assumption for the sample and medium, which is generally valid in colorless transparent materials. This assumption forces the scattering potential $\chi(x,y,z)$ to be wavelength-independent.

*3.2 Volumetric k-space and axial resolution*

Since each wavelength yields different $k = |\mathbf{k}| = |\mathbf{k}^{(i)}|$, the corresponding value of $\tilde{\chi}\left(\mathbf{k} - \mathbf{k}^{(i)}\right)$ in Eq. (4) represents spherical shells with different radii in 3-D $k$-space (Fig. 3). Therefore, unlike in monochromatic cases, the associated $\tilde{\chi}(k_x, k_y, k_z)$ exhibits a certain axial thickness $\Delta k_z$ in $k$-space that governs finite axial image resolution [57, 58]. Such an incoherence-based axial sectioning ability (i.e., coherence gating) also forms the basis of optical coherence tomography (OCT), which shares the fundamental physics with ODT, namely coherent diffraction of light [69].

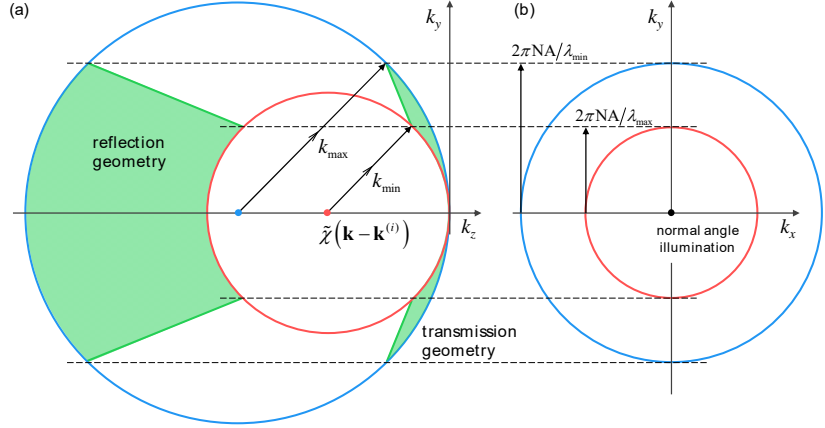

Fig. 3. *k*-space volume coverage of $\tilde{\chi}(k_x, k_y, k_z)$ during incoherent illumination (normal angle) over the sample scattering potential. (a) the $k_y k_z$ and (b) $k_x k_y$ planes are shown. Transmission and reflection geometries are compared. To visualize the effect of broadband illumination, we set $\lambda_{min}$ = 400 nm, $\lambda_{max}$ = 700 nm, *m* =1, and NA = 0.7 in the figure.

In general, the *k*-space volume cannot be acquired in a single measurement. Therefore, the axial scanning method has been a common solution, as in previous investigations into incoherent ODT [57, 58] or in time-domain OCT setups [69]. However, unlike in reflective geometries, transmission geometry exhibits far thinner axial thickness $\Delta k_z$ even for highly incoherent illuminations, which directly presents the far poorer axial resolution and sectioning abilities (Fig. 3). Furthermore, $\Delta k_z$ rapidly diminishes as the lateral frequency $k_\perp = \sqrt{k^2 - k_z^2}$ decreases, and eventually becomes zero for $k_\perp = 0$ (Fig. 3). This issue is similar to the missing cone problem in ODT [28] but far more severe; and is especially critical for weakly scattering samples such as biological cells, where the scattering information is mostly distributed near the origin. Therefore, for the majority of biological applications, we find incoherence-induced $\Delta k_z$ barely provides more than two distinguishable points along the *z*-axis. This is identical to assuming uniform $\tilde{\chi}(\mathbf{k} - \mathbf{k}^{(i)})$ along the $k_z$ axis in Eq. (4) and Fig 3.

Notice that the decoherence effect originating from the sample-induced OPL is not related to the axial resolution of the system. For example, a thick dielectric slab would induce the decoherence regardless of its axial position, which directly shows the axial indistinguishability of the system.

### 3.3 Weight function

According to the discussions in the previous subsections, $\tilde{\chi}(\mathbf{k} - \mathbf{k}^{(i)})$ could be assumed to be independent to $k_z$, and Eq. (4) can be simplified as

$$\tilde{U}^{(s)}(k_x, k_y) = \frac{i}{2} w(k_\perp; P) \tilde{\chi}(\mathbf{k} - \mathbf{k}^{(i)}), \text{ where} \qquad (5)$$

$$w(k_\perp; P) = \int_{k_\perp}^{\infty} \frac{k^2}{\sqrt{k^2 - k_\perp^2}} \frac{\partial P}{\partial k} dk \qquad (6)$$

is a weight function related to the lateral frequency $k_\perp$ and the normalized power spectral density $\partial P / \partial k$, and $k_{\underline{\perp}} = (m/\text{NA}) k_\perp$ is the minimum transferrable *k* for a given $k_\perp$ value and NA of the objective lens (Fig. 4).

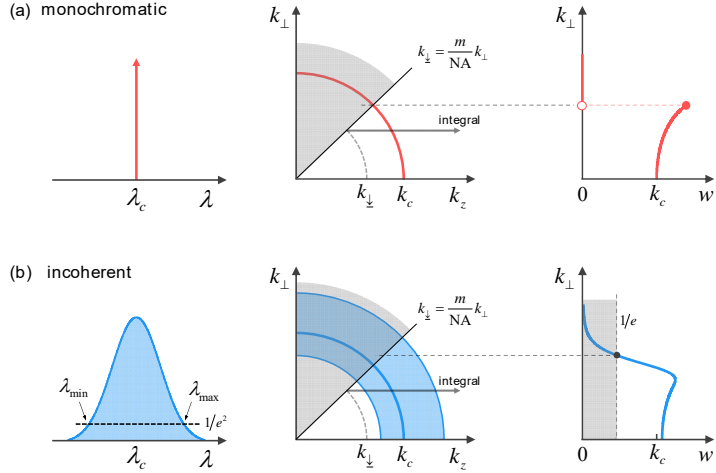

Fig. 4. Weight functions in (a) monochromatic illumination, and (b) incoherent illumination. The center wavelength is identical in both situations. To visualize the effect of broadband illumination, we set $\lambda_{min}$ = 400 nm, $\lambda_c$ = 550 nm, $\lambda_{min}$ = 700 nm, $m$ = 1, and NA = 0.7.

Therefore, the sample scattering potential $\tilde{\chi}(\mathbf{k} - \mathbf{k}^{(i)})$ could be reconstructed from a single incoherent measurement $\tilde{U}^{(s)}(k_x, k_y)$ by calculating $w(k_\perp; P)$.

As shown in Fig. 4, $w(k_\perp; P)$ for monochromatic cases increases with $k_\perp$. This is the inverse cosine factor originating from the curvature of a spherical shell. The $w(k_\perp; P)$ exhibits similar trend for incoherent cases and low $k_\perp$ values. However, $w(k_\perp; P)$ rapidly decreases as $k_\perp$ excludes the lower $k$ (i.e., longer $\lambda$) portion of given spectrum. In fact, this feature significantly lowers the signal-to-noise ratio for higher $k_\perp$ values.

In practical situations, we need to remove the residuals from the spectrum to prevent noise amplification when dividing by $w(k_\perp; P) \approx 0$. Therefore, we set the $1/e$ of the maximum weight as the maximum attainable $k_\perp$ value [Fig. 4(b)]. Similarly, we bound the valid spectral domain to define the achievable solid volume of $\tilde{\chi}(\mathbf{k} - \mathbf{k}^{(i)})$ (Fig. 5), where $k_{min} = 2\pi m/\lambda_{max}$ and $k_{max} = 2\pi m/\lambda_{min}$ are the wavenumbers at a factor $1/e^2$ of the peak spectral density [Fig. 4(b)].

### 3.4 Angle scanning illumination

Although we use the term "angle scanning" to prevent confusion, it is not the exact description because we utilize DMD-based diffractive modulation that conserves the scanning lateral frequency $k_\perp = k \sin\theta$ rather than varying the scanning angle $\theta$. In other words, each wavelength is incident at a different angle, $\theta = \sin^{-1}(k_\perp / k)$ during a single illumination [Fig. 2(b))].

However, longer wavelengths would be filtered out by the condenser lens as the lateral scanning frequency $k_\perp$ increases, which should be avoided to prevent intensity loss and interference visibility. Therefore, the maximum lateral scanning frequency $k_\perp$ should be limited by $(\mathrm{NA}^{(i)}/m) k_{min}$ [Fig. 5(b)]. Accordingly, the effective illumination NA becomes a function of wavelength $\mathrm{NA}^{(i)}(k_{min}/k)$, which is always smaller than the given $\mathrm{NA}^{(i)}$ of the condenser.

We compare volumetric coverage of incoherent angle scanning with monochromatic scanning with the same $\lambda_c$ value in identical optical setups (Fig. 5). We set NA= $\mathrm{NA}^{(i)}$ in this comparison for simplicity. We find incoherent illumination in the circular scanning case has a significant advantage. The decreased illumination angle at shorter wavelengths effectively covers the $\tilde{\chi}(k_z < 0)$ region, which cannot be acquired in monochromatic circular scanning with maximum scanning radius [Fig. 5(c)]. However, such advantage relatively fades in full aperture scanning since a small volume could also be sampled from circular scanning with incoherent illumination [Fig. 5(d)].

It is noteworthy that similar *k*-space volume information could be acquired with spatially incoherent illumination and axial scanning [48, 57, 58], because it could be regarded as an incoherent summation of different plane waves. Indeed, a proper weight function should be accompanied to quantify $\tilde{\chi}(k_x, k_y, k_z)$ from measurements.

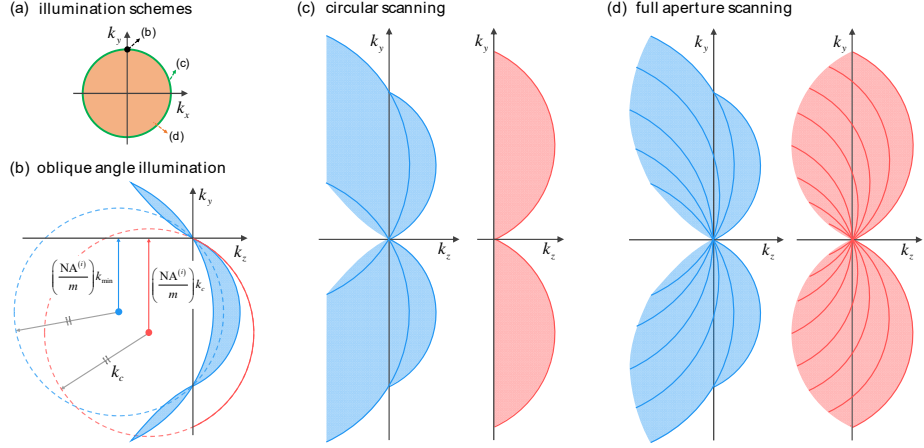

Fig. 5. Angle-scanning illumination. (a) *k*-space visualization of illumination schemes. *k*-space volume coverage of $\tilde{\chi}(k_x, k_y, k_z)$ in (b) single oblique angle illumination, (c) circular scanning, and (d) full aperture scanning cases. All graphs are depicted in a $k_x = 0$ slice to emphasize the axial coverage. The same parameters as in Fig. 4(b) were applied in this figure.

## 4. Results and discussions

### 4.1 On the proper degree of incoherency

According to the theoretical analysis in Section 3, we find the illumination bandwidth should be chosen carefully. Unlike our expectation in Section 1, too broad a bandwidth also raises several fundamental and practical disadvantages, such as group delay dispersion mismatch, sample dispersion effect, and illumination angle limitation. Meanwhile, the optical sectioning ability is not as significant as in reflection geometries. Therefore, we find that a minimum degree of incoherence is preferred in order to achieve interference suppression induced by unwanted beam paths. In many ODT setups, for instance, the coverslip induces multiple reflections that exhibit minimum additional OPL, which is on the order of 100 μm. In such cases, a coherence length of $l_c \sim 10$ μm could safely meet the noise suppression goal, which corresponds to ~10 nm spectral bandwidth in the visible range. In the light of above discussion, we utilize a bandpass filter to limit the spectral bandwidth to 17 nm [Fig. 1(a)]. The corresponding weight function is shown in Fig. 6(c), which has a similar shape as the previous expectation [Fig. 4(b)], but steeper slope due to far narrower bandwidth. As a low-priced alternative, we suggest a superluminescent diodes (SLD), which provides low-coherent light of ~10 nm spectral bandwidth with good spatial coherency.

### 4.2 Optical diffraction tomography results

Thanks to the diffractive modulation, full-field interference of incoherent illumination is steadily maintained during the entire scanning procedure [Fig. 6(a)]. The optical field $U(x, y)$ could be retrieved with a conventional off-axis modulation scheme from the acquired interferograms [Fig. 6(b)].

As discussed in Section 3.3, the sampling area radius in the plane is set as a factor $1/e$ of the maximum weight [Fig. 6(c)]. The incident field $U^{(i)}(x, y)$ is measured using an identical procedure without the sample, and the scattered field $U^{(s)}(x, y)$ can be calculated from Eq. (2) [Figs. 6(d)-6(f)]. Notice, we utilize the first Rytov approximation rather than using Eq. (2) directly (i.e., the first Born approximation) to achieve more reliable results in micro-sized samples [67, 68]. Substituting $\tilde{U}^{(s)}(k_x, k_y)$ and $w(k_\perp; P)$ into Eq. (5), we obtain $\tilde{\chi}(\mathbf{k} - \mathbf{k}^{(i)})$, which is directly related to the 3-D RI distribution $n(x, y, z)$ [Fig. 6(g)]. In order to manage the inevitable missing cone issue in ODT with transmission geometry, we applied edge-preserving regularization as discussed in Ref. [28].

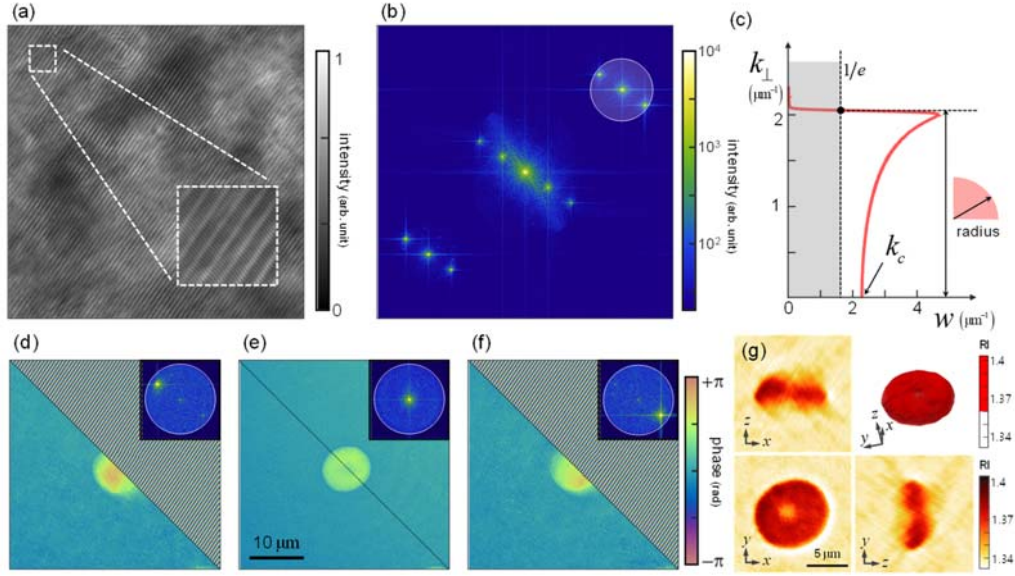

Fig. 6. Experimental demonstrations. (a) Interferogram measured with incoherent illumination. (b) Panel (a) in *k*-space. The off-axis sampling area is indicated with a black-lined circle, whose radius is defined by (c) the weight function. (d-f) The retrieved sample phase maps corresponding to each peak in (b). Phase ramp is compensated for half of the images to visualize the sample information. (g-j) Corresponding reconstructed 3-D RI tomogram.

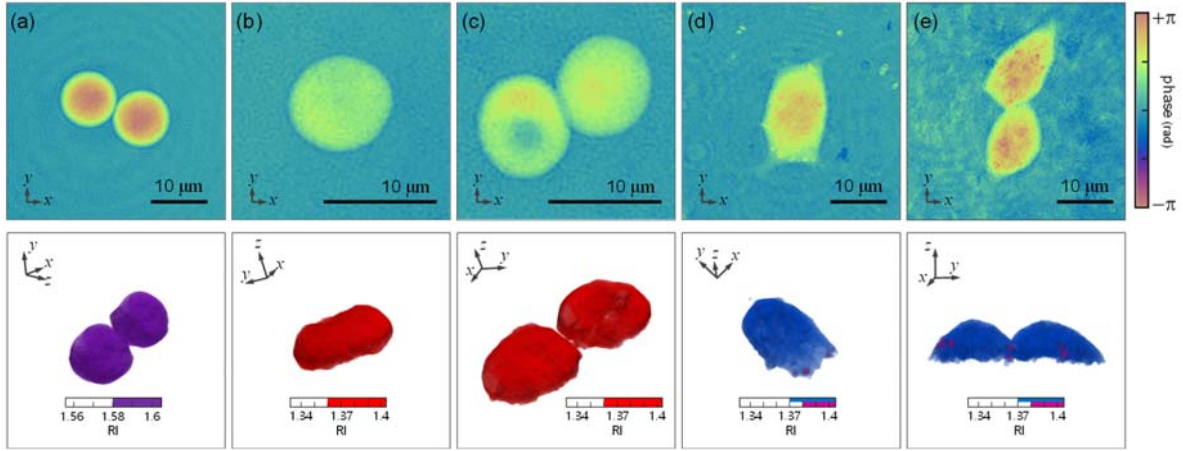

Fig. 7. Experimental normal angle phase images (upper) and corresponding 3-D RI tomogram results (lower) in various samples. (a) Polystyrene microsphere with 10 μm diameter immersed in RI matching oil. (b-c) Red blood cells diluted with Alservier's solution. (d-e) Rat pheochromocytoma (PC-12) cells cultured for 24 hours.

To test the versatility of the proposed method, we measured the 3-D RI distribution of various samples (Fig. 7). Polystyrene microspheres with 10 μm diameter were prepared using proper RI matching by using $m = 1.561$ immersion oil [Fig. 7(a)]. Red blood cells donated from a healthy donor were prepared using Alservier's solution as a diluent (A3551, Sigma-Aldrich Inc.) [Figs. 7(b) and 7(c)]. Rat pheochromocytoma (PC-12) cells were cultured in an incubator for 24 hours at 37 °C and 5% $CO_2$ concentration [Figs. 7(d) and 7(e)]. Despite the use of low-coherent light, we find subtle diffraction noise coupled with the sample information. For example, the concentric fringe observed in Fig. 7(a) could originate from a stain on the coverslips. Notice that such low-angle diffraction cannot be decoupled via incoherent illumination due to the poor axial resolution in transmission geometry (see Section 3.2). For the cultured cells, the inhomogeneous background is mainly originated from the unwanted cell debris or microorganisms [Fig. 7(e)].

## 5. Conclusion

In this article, we present a general off-axis angle-scanning incoherent ODT technique based on the use of an MZ interferometer. Diffraction tilting was used to maintain interference during the scanning procedure. As a controllable diffractive unit, a DMD was used to maximize the acquisition speed. Chromatic dispersion induced from the blaze angle of the DMD pixels is compensated by an identical DMD in the idle state.

We also review the principle of incoherent ODT reconstruction and clarify its difference from conventional coherent cases by introducing spectrum-based weight functions. Several assumptions were introduced and discussed during the reconstruction procedure. In addition to such inevitable assumptions, we find several fundamental and practical limitations of incoherent illumination that effectively reduce the ODT volumetric sampling capacity, which dilutes the advantages of incoherent illumination. Taking such disadvantages into account, we discuss and suggest the proper degree of incoherency for the maximum quality of ODT. Based on the proposed setup and reconstruction principle, we successfully performed low-coherent ODT reconstruction in various samples.

In view of the use of incoherent light sources for 3D label-free imaging of biological samples, it could make the translation of ODT to biological and medical applications. In addition, the approach demonstrated here is general and can also be combined with other imaging modalities, including birefringence [70, 71], spectroscopic [72, 73], and fluorescence signals [74-76]. Going forward, we envision that, by full exploitation of low-coherent light sources, QPI could expand its applications.

## Acknowledgement

Dr. K. Lee, Mr. S. Shin, and Prof. Park have financial interests in Tomocube Inc., a company that commercializes ODT and QPI instruments and is one of the sponsors of the work.

## Funding

This work was supported by KAIST, BK21+ program, Tomocube, and National Research Foundation of Korea (2015R1A3A2066550, 2017M3C1A3013923, 2014K1A3A1A09063027).

## References and links

1. G. Popescu, *Quantitative Phase Imaging of Cells and Tissues* (McGraw-Hill Education, 2011).
2. K. Lee, K. Kim, J. Jung, J. Heo, S. Cho, S. Lee, G. Chang, Y. Jo, H. Park, and Y. Park, "Quantitative Phase Imaging Techniques for the Study of Cell Pathophysiology: From Principles to Applications," Sensors **13**, 4170 (2013).
3. B. Kemper and G. von Bally, "Digital holographic microscopy for live cell applications and technical inspection," Appl. Opt. **47**, A52-A61 (2008).
4. M. Mir, Z. Wang, Z. Shen, M. Bednarz, R. Bashir, I. Golding, S. G. Prasanth, and G. Popescu, "Optical measurement of cycle-dependent cell growth," Proceedings of the National Academy of Sciences **108**, 13124 (2011).
5. M. Mir, A. Bergamaschi, B. S. Katzenellenbogen, and G. Popescu, "Highly Sensitive Quantitative Imaging for Monitoring Single Cancer Cell Growth Kinetics and Drug Response," PLOS ONE **9**, e89000 (2014).
6. H. Ding, Z. Wang, F. Nguyen, S. A. Boppart, and G. Popescu, "Fourier transform light scattering of inhomogeneous and dynamic structures," Physical Review Letters **101**, 238102 (2008).
7. J. Jung and Y. Park, "Spectro-angular light scattering measurements of individual microscopic objects," Opt. Express **22**, 4108-4114 (2014).
8. Y. Jo, J. Jung, M.-h. Kim, H. Park, S.-J. Kang, and Y. Park, "Label-free identification of individual bacteria using Fourier transform light scattering," Opt. Express **23**, 15792-15805 (2015).
9. Z. Wang, H. Ding, and G. Popescu, "Scattering-phase theorem," Opt. Lett. **36**, 1215-1217 (2011).
10. M. Lee, E. Lee, J. Jung, H. Yu, K. Kim, J. Yoon, S. Lee, Y. Jeong, and Y. Park, "Label-free optical quantification of structural alterations in Alzheimer's disease," Scientific reports **6**, 31034 (2016).
11. Y. Kim, J. Jeong, J. Jang, M. W. Kim, and Y. Park, "Polarization holographic microscopy for extracting spatio-temporally resolved Jones matrix," Opt. Express **20**, 9948-9955 (2012).
12. J. Jung, J. Kim, M.-K. Seo, and Y. Park, "Measurements of polarization-dependent angle-resolved light scattering from individual microscopic samples using Fourier transform light scattering," Opt. Express **26**, 7701-7711 (2018).
13. S. A. Alexandrov, T. R. Hillman, T. Gutzler, and D. D. Sampson, "Synthetic Aperture Fourier Holographic Optical Microscopy," Physical Review Letters **97**, 168102 (2006).
14. T. R. Hillman, T. Gutzler, S. A. Alexandrov, and D. D. Sampson, "High-resolution, wide-field object reconstruction with synthetic aperture Fourier holographic optical microscopy," Opt. Express **17**, 7873-7892 (2009).
15. K. Lee, H.-D. Kim, K. Kim, Y. Kim, T. R. Hillman, B. Min, and Y. Park, "Synthetic Fourier transform light scattering," Opt. Express **21**, 22453-22463 (2013).
16. Y. Sung, W. Choi, C. Fang-Yen, K. Badizadegan, R. R. Dasari, and M. S. Feld, "Optical diffraction tomography for high resolution live cell imaging," Opt. Express **17**, 266-277 (2009).
17. R. Barer, "Interference microscopy and mass determination," Nature **169**, 366-367 (1952).
18. K. Kim, J. Yoon, S. Shin, S. Lee, S.-A. Yang, and Y. Park, "Optical diffraction tomography techniques for the study of cell pathophysiology," Journal of Biomedical Photonics & Engineering **2**(2016).
19. E. Wolf, "Three-dimensional structure determination of semi-transparent objects from holographic data," Optics Communications **1**, 153-156 (1969).
20. V. Lauer, "New approach to optical diffraction tomography yielding a vector equation of diffraction tomography and a novel tomographic microscope," Journal of Microscopy **205**, 165-176 (2002).
21. Y. Sung, N. Lue, B. Hamza, J. Martel, D. Irimia, R. R. Dasari, W. Choi, Z. Yaqoob, and P. So, "Three-Dimensional Holographic Refractive-Index Measurement of Continuously Flowing Cells in a Microfluidic Channel," Physical Review Applied **1**, 014002 (2014).
22. P. Hosseini, Y. Sung, Y. Choi, N. Lue, Z. Yaqoob, and P. So, "Scanning color optical tomography (SCOT)," Opt. Express **23**, 19752-19762 (2015).

23. K. Kim, J. Yoon, S. Shin, S. Lee, S.-A. Yang, and Y. Park, "Optical diffraction tomography techniques for the study of cell pathophysiology," Journal of Biomedical Photonics & Engineering **2**, 020201 (2016).
24. F. Charrière, A. Marian, F. Montfort, J. Kuehn, T. Colomb, E. Cuche, P. Marquet, and C. Depeursinge, "Cell refractive index tomography by digital holographic microscopy," Opt. Lett. **31**, 178-180 (2006).
25. M. Habaza, B. Gilboa, Y. Roichman, and N. T. Shaked, "Tomographic phase microscopy with 180 rotation of live cells in suspension by holographic optical tweezers," Opt. Lett. **40**, 1881-1884 (2015).
26. P. Müller, M. Schürmann, C. J. Chan, and J. Guck, "Single-cell diffraction tomography with optofluidic rotation about a tilted axis," in *Optical Trapping and Optical Micromanipulation XII*, (International Society for Optics and Photonics, 2015), 95480U.
27. F. Merola, P. Memmolo, L. Miccio, R. Savoia, M. Mugnano, A. Fontana, G. D'ippolito, A. Sardo, A. Iolascon, and A. Gambale, "Tomographic flow cytometry by digital holography," Light: Science & Applications **6**, e16241 (2017).
28. J. Lim, K. Lee, K. H. Jin, S. Shin, S. Lee, Y. Park, and J. C. Ye, "Comparative study of iterative reconstruction algorithms for missing cone problems in optical diffraction tomography," Opt. Express **23**, 16933-16948 (2015).
29. J. Hur, K. Kim, S. Lee, H. Park, and Y. Park, "Melittin-induced alterations in morphology and deformability of human red blood cells using quantitative phase imaging techniques," Scientific reports **7**, 9306 (2017).
30. S. Lee, H. Park, K. Kim, Y. Sohn, S. Jang, and Y. Park, "Refractive index tomograms and dynamic membrane fluctuations of red blood cells from patients with diabetes mellitus," Scientific reports **7**, 1039 (2017).
31. J. Yoon, Y. Jo, M.-h. Kim, K. Kim, S. Lee, S.-J. Kang, and Y. Park, "Identification of non-activated lymphocytes using three-dimensional refractive index tomography and machine learning," Scientific Reports **7**, 6654 (2017).
32. S. Kwon, Y. Lee, Y. Jung, J. H. Kim, B. Baek, B. Lim, J. Lee, I. Kim, and J. Lee, "Mitochondria-targeting indolizino [3, 2-c] quinolines as novel class of photosensitizers for photodynamic anticancer activity," European journal of medicinal chemistry **148**, 116-127 (2018).
33. T. H. Nguyen, M. E. Kandel, M. Rubessa, M. B. Wheeler, and G. Popescu, "Gradient light interference microscopy for 3D imaging of unlabeled specimens," Nature communications **8**, 210 (2017).
34. J. Oh, G.-H. Lee, J. Noh, S. Shin, B. J. Lee, Y. Nam, and Y. Park, "Optical measurements of three-dimensional microscopic temperature distributions around gold nanorods excited by surface plasmonics," arXiv preprint arXiv:1804.00758 (2018).
35. P. Marquet, B. Rappaz, P. J. Magistretti, E. Cuche, Y. Emery, T. Colomb, and C. Depeursinge, "Digital holographic microscopy: a noninvasive contrast imaging technique allowing quantitative visualization of living cells with subwavelength axial accuracy," Opt. Lett. **30**, 468-470 (2005).
36. G. Popescu, T. Ikeda, R. R. Dasari, and M. S. Feld, "Diffraction phase microscopy for quantifying cell structure and dynamics," Opt. Lett. **31**, 775-777 (2006).
37. Y. Park, G. Popescu, K. Badizadegan, R. R. Dasari, and M. S. Feld, "Diffraction phase and fluorescence microscopy," Opt. Express **14**, 8263-8268 (2006).
38. C. Fang-Yen, S. Oh, Y. Park, W. Choi, S. Song, H. S. Seung, R. R. Dasari, and M. S. Feld, "Imaging voltage-dependent cell motions with heterodyne Mach-Zehnder phase microscopy," Opt. Lett. **32**, 1572-1574 (2007).
39. S. Shin, K. Kim, K. Lee, S. Lee, and Y. Park, "Effects of spatiotemporal coherence on interferometric microscopy," Opt. Express **25**, 8085-8097 (2017).
40. P. Ferraro, S. De Nicola, A. Finizio, G. Coppola, S. Grilli, C. Magro, and G. Pierattini, "Compensation of the inherent wave front curvature in digital holographic coherent microscopy for quantitative phase-contrast imaging," Appl. Opt. **42**, 1938-1946 (2003).
41. T. Colomb, F. Montfort, J. Kühn, N. Aspert, E. Cuche, A. Marian, F. Charrière, S. Bourquin, P. Marquet, and C. Depeursinge, "Numerical parametric lens for shifting, magnification, and complete aberration compensation in digital holographic microscopy," J. Opt. Soc. Am. A **23**, 3177-3190 (2006).
42. L. Miccio, D. Alfieri, S. Grilli, P. Ferraro, A. Finizio, L. D. Petrocellis, and S. D. Nicola, "Direct full compensation of the aberrations in quantitative phase microscopy of thin objects by a single digital hologram," Applied Physics Letters **90**, 041104 (2007).
43. I. Choi, K. Lee, and Y. Park, "Compensation of aberration in quantitative phase imaging using lateral shifting and spiral phase integration," Opt. Express **25**, 30771-30779 (2017).
44. Y. Park, W. Choi, Z. Yaqoob, R. Dasari, K. Badizadegan, and M. S. Feld, "Speckle-field digital holographic microscopy," Opt. Express **17**, 12285-12292 (2009).
45. F. Dubois, M.-L. N. Requena, C. Minetti, O. Monnom, and E. Istasse, "Partial spatial coherence effects in digital holographic microscopy with a laser source," Appl. Opt. **43**, 1131-1139 (2004).
46. H. Farrokhi, J. Boonruangkan, B. J. Chun, T. M. Rohith, A. Mishra, H. T. Toh, H. S. Yoon, and Y.-J. Kim, "Speckle reduction in quantitative phase imaging by generating spatially incoherent laser field at electroactive optical diffusers," Opt. Express **25**, 10791-10800 (2017).
47. X. Li, T. Yamauchi, H. Iwai, Y. Yamashita, H. Zhang, and T. Hiruma, "Full-field quantitative phase imaging by white-light interferometry with active phase stabilization and its application to biological samples," Opt. Lett. **31**, 1830-1832 (2006).
48. Y. Choi, T. D. Yang, K. J. Lee, and W. Choi, "Full-field and single-shot quantitative phase microscopy using dynamic speckle illumination," Opt. Lett. **36**, 2465-2467 (2011).
49. Y. Choi, P. Hosseini, W. Choi, R. R. Dasari, P. T. C. So, and Z. Yaqoob, "Dynamic speckle illumination wide-field reflection phase microscopy," Opt. Lett. **39**, 6062-6065 (2014).
50. Z. Wang, L. Millet, M. Mir, H. Ding, S. Unarunotai, J. Rogers, M. U. Gillette, and G. Popescu, "Spatial light interference microscopy (SLIM)," Opt. Express **19**, 1016-1026 (2011).
51. B. Bhaduri, H. Pham, M. Mir, and G. Popescu, "Diffraction phase microscopy with white light," Opt. Lett. **37**, 1094-1096 (2012).
52. Y. Baek, K. Lee, J. Yoon, K. Kim, and Y. Park, "White-light quantitative phase imaging unit," Opt. Express **24**, 9308-9315 (2016).
53. T. H. Nguyen, C. Edwards, L. L. Goddard, and G. Popescu, "Quantitative phase imaging with partially coherent illumination," Opt. Lett. **39**, 5511-5514 (2014).
54. P. Bon, S. Monneret, and B. Wattellier, "Noniterative boundary-artifact-free wavefront reconstruction from its derivatives," Appl. Opt. **51**, 5698-5704 (2012).
55. A. Descloux, K. Grußmayer, E. Bostan, T. Lukes, A. Bouwens, A. Sharipov, S. Geissbuehler, A.-L. Mahul-Mellier, H. Lashuel, and M. Leutenegger, "Combined multi-plane phase retrieval and super-resolution optical fluctuation imaging for 4D cell microscopy," Nature Photonics **12**, 165 (2018).
56. M. Rinehart, Y. Zhu, and A. Wax, "Quantitative phase spectroscopy," Biomed. Opt. Express **3**, 958-965 (2012).
57. T. Kim, R. Zhou, M. Mir, S. D. Babacan, P. S. Carney, L. L. Goddard, and G. Popescu, "White-light diffraction tomography of unlabelled live cells," Nature Photonics **8**, 256 (2014).
58. P. Bon, S. Aknoun, S. Monneret, and B. Wattellier, "Enhanced 3D spatial resolution in quantitative phase microscopy using spatially incoherent illumination," Opt. Express **22**, 8654-8671 (2014).
59. S. Chowdhury, W. J. Eldridge, A. Wax, and J. Izatt, "Refractive index tomography with structured illumination," Optica **4**, 537-545 (2017).

60. R. Zhou, T. Kim, L. L. Goddard, and G. Popescu, "Inverse scattering solutions using low-coherence light," Opt. Lett. **39**, 4494-4497 (2014).
61. M. Takeda, H. Ina, and S. Kobayashi, "Fourier-transform method of fringe-pattern analysis for computer-based topography and interferometry," J. Opt. Soc. Am. **72**, 156-160 (1982).
62. Z. Yaqoob, T. Yamauchi, W. Choi, D. Fu, R. R. Dasari, and M. S. Feld, "Single-shot Full-field reflection phase microscopy," Opt. Express **19**, 7587-7595 (2011).
63. S. Chowdhury, W. J. Eldridge, A. Wax, and J. A. Izatt, "Structured illumination microscopy for dual-modality 3D sub-diffraction resolution fluorescence and refractive-index reconstruction," Biomed. Opt. Express **8**, 5776-5793 (2017).
64. X. Chen, B.-b. Yan, F.-j. Song, Y.-q. Wang, F. Xiao, and K. Alameh, "Diffraction of digital micromirror device gratings and its effect on properties of tunable fiber lasers," Appl. Opt. **51**, 7214-7220 (2012).
65. S. Shin, K. Kim, J. Yoon, and Y. Park, "Active illumination using a digital micromirror device for quantitative phase imaging," Opt. Lett. **40**, 5407-5410 (2015).
66. K. Lee, K. Kim, G. Kim, S. Shin, and Y. Park, "Time-multiplexed structured illumination using a DMD for optical diffraction tomography," Opt. Lett. **42**, 999-1002 (2017).
67. A. C. Kak and M. Slaney, *Principles of Computerized Tomographic Imaging* (Society for Industrial and Applied Mathematics, 2001).
68. A. J. Devaney, *Mathematical Foundations of Imaging, Tomography and Wavefield Inversion* (Cambridge University Press, 2012).
69. D. Huang, E. Swanson, C. Lin, J. Schuman, W. Stinson, W. Chang, M. Hee, T. Flotte, K. Gregory, C. Puliafito, and a. et, "Optical coherence tomography," Science **254**, 1178-1181 (1991).
70. Z. Wang, L. J. Millet, M. U. Gillette, and G. Popescu, "Jones phase microscopy of transparent and anisotropic samples," Opt. Lett. **33**, 1270-1272 (2008).
71. Y. Kim, J. Jeong, J. Jang, M. W. Kim, and Y. Park, "Polarization holographic microscopy for extracting spatio-temporally resolved Jones matrix," Opt. Express **20**, 9948-9955 (2012).
72. J. Jung, K. Kim, J. Yoon, and Y. Park, "Hyperspectral optical diffraction tomography," Opt. Express **24**, 2006-2012 (2016).
73. A. Ojaghi, M. E. Fay, W. A. Lam, and F. E. Robles, "Ultraviolet Hyperspectral Interferometric Microscopy," Scientific reports **8**, 9913 (2018).
74. S. Chowdhury, W. J. Eldridge, A. Wax, and J. A. Izatt, "Structured illumination multimodal 3D-resolved quantitative phase and fluorescence sub-diffraction microscopy," Biomed. Opt. Express **8**, 2496-2518 (2017).
75. K. Kim, W. S. Park, S. Na, S. Kim, T. Kim, W. Do Heo, and Y. Park, "Correlative three-dimensional fluorescence and refractive index tomography: bridging the gap between molecular specificity and quantitative bioimaging," Biomed. Opt. Express **8**, 5688-5697 (2017).
76. M. Schürmann, G. Cojoc, S. Girardo, E. Ulbricht, J. Guck, and P. Müller, "Three-dimensional correlative single-cell imaging utilizing fluorescence and refractive index tomography," Journal of biophotonics **11**, e201700145 (2018).